\documentclass[sigconf,natbib=false,anonymous=false]{acmart}

\AtBeginDocument{%
  }

\renewcommand\footnotetextcopyrightpermission[1]{}

\RequirePackage[
  datamodel=acmdatamodel,
  style=acmauthoryear,
  ]{biblatex}

\usepackage{float}
\usepackage{adjustbox}
\usepackage{tikz}
\usepackage{booktabs}
\usepackage{subcaption}
\usepackage{tabularx}
\usetikzlibrary{shapes.geometric, arrows, positioning, decorations.pathreplacing, decorations.text}
\usepackage{hyperref}
\begin{document}

\title[InPars+: Supercharging Synthetic Data Generation for Information Retrieval Systems]{InPars+: Supercharging Synthetic Data Generation \\ for Information Retrieval Systems}

\author{Matey Krastev}
\authornotemark[1]
\affiliation{%
  \institution{University of Amsterdam}
  \city{Amsterdam}
  \country{the Netherlands}
}

\author{Miklos Hamar}
\authornotemark[1]
\affiliation{%
  \institution{University of Amsterdam}
  \city{Amsterdam}
  \country{the Netherlands}
}

\author{Danilo Toapanta}
\authornotemark[1]
\affiliation{%
  \institution{University of Amsterdam}
  \city{Amsterdam}
  \country{the Netherlands}
}

\author{Jesse Brouwers}
\authornotemark[1]
\affiliation{%
  \institution{University of Amsterdam}
  \city{Amsterdam}
  \country{the Netherlands}
}
\author{Yibin Lei}
\authornotemark[1]
\affiliation{%
  \institution{University of Amsterdam}
  \city{Amsterdam}
  \country{the Netherlands}
}

\begin{abstract}
This work revisits and extends synthetic query generation pipelines for Neural Information Retrieval (NIR) by leveraging the InPars Toolkit, a reproducible, end-to-end framework for generating training data using large language models (LLMs). We first assess the reproducibility of the original InPars, InPars-V2, and Promptagator pipelines on the SciFact benchmark and validate their effectiveness using open-source reranker and generator models. Building on this foundation, we introduce two key extensions to the pipeline: (1) fine-tuning a query generator LLM via Contrastive Preference Optimization (CPO) to improve the signal quality in generated queries, and (2) replacing static prompt templates with dynamic, Chain-of-Thought (CoT) optimized prompts using the DSPy framework. Our results show that both extensions reduce the need for aggressive filtering while improving retrieval performance. All code, models, and synthetic datasets are publicly released to support further research: \href{https://github.com/danilotpnta/IR2-project}{https://github.com/danilotpnta/IR2-project}.
\end{abstract}


\maketitle
\pagestyle{plain}

\section{Introduction}

%
%
The training of Neural Information Retrieval (NIR) models requires a substantial amount of annotated data. Typically, a dataset is a collection of documents, paired with queries and human-labelled relevance judgments that connect the two. These relevance judgements, however, are hard and costly to acquire. For example, a human annotator typically requires at least one minute on average to assess the relevance of a query-document pair \cite{boytsov2023inpars}.

To address this challenge, prior research has explored leveraging multi-billion-parameter Large Language Models (LLMs) to generate relevant queries synthetically. Notable examples include InPars \cite{bonifacio2022inpars}, its extensions InPars-V2 \cite{jeronymo2023inpars} and InPars-light \cite{boytsov2023inpars}, as well as Promptagator \cite{dai2022promptagator}. These approaches are commonly referred to as QGen pipelines\cite{chaudhary2023itsrelativesynthetic}, where a document (re-)ranking model, such as MonoT5 \cite{nogueira2020document}, is fine-tuned using the synthetic data to improve downstream retrieval performance. These pipelines show promising results, which motivate further experimentation in this direction.

While these methods offer researchers flexible pipelines for enhancing Neural Information Retrieval (NIR) models in data-scarce scenarios, reproducing and utilising the aforementioned QGen pipelines remains a challenging task. To address this issue, Abonizio et al. introduced InPars Toolkit~\cite{abonizio2023inpars}, an open-source end-to-end pipeline for synthetic data generation with GPU support. The toolkit enables researchers and practitioners to efficiently experiment with and implement IR pipelines, significantly reducing the initial implementation complexity.

In the first part of this study, we aim to reproduce their results. Specifically, we aim to validate the following claims:

\begin{enumerate}
    \item InPars Toolkit is an end-to-end \textit{reproducible pipeline} for synthetic data generation.
    \item It provides a comprehensive guideline for reproducing in full InPars-V1 \cite{bonifacio2022inpars}, InPars-V2 \cite{jeronymo2023inpars}, as well as in part Promptagator \cite{dai2022promptagator}.
    \item The toolkit has \textit{plug-and-play} functionality, allowing for seamless integration of alternative LLMs.
\end{enumerate}

In doing so, we aim to validate the transparency and reliability of adopting and extending the InPars Toolkit for future research.

Secondly, a common key insight from these studies is that the synthetically generated data need certain filtering mechanisms to ensure high-quality training data for the downstream model. In our study, we found that this filtering step is computationally expensive and can result in a significant amount of generated data being discarded~\cite{bonifacio2022inpars, dai2022promptagator}. 

To address this, we propose two main extensions to the InPars Toolkit: (1) fine-tuning the generator model using Contrastive Preference Optimization (CPO)~\cite{xu2024contrastive} to improve the quality of generated queries, and (2) employing dynamic prompt optimization using DSPy~\cite{khattab2024dspy} to enhance the prompt templates used in the data generation process.

By fine-tuning the generator model with CPO, we aim to reduce the noise in the generated queries, thereby increasing the proportion of high-quality queries and minimizing the need for extensive filtering. This approach leverages a preference-based optimization technique to guide the generator model towards producing more relevant and useful queries.

Additionally, by integrating DSPy for dynamic prompt optimization, we aim to replace the static prompt templates currently used in the InPars Toolkit with more adaptive and contextually appropriate prompts. This method utilizes the LLM's own capabilities to determine what constitutes a relevant query, potentially improving the overall quality of the generated data and further reducing the reliance on filtering.


Our contributions can be summarised as follows:
\begin{itemize}
    \item We validate the main claims of \cite{abonizio2023inpars} and the InPars Toolkit.
    \item We use a Llama3.1-8B model to generate synthetic data for Scifact, and finetune MonoT5-3B reranker models on the generated sets of synthetic data.
    \item We extend the InPars Toolkit with a query generator fine-tuning stage, where we finetune an LLM on a subset of MS-Marco \cite{bajaj2016ms} using a CPO training scheme \cite{xu2024contrastive}.
    \item We explore the application of DSPy in the synthetic data generation pipeline, aiming to improve upon the currently used static prompt templates.
    \item We make the synthetic data and models generated in this study publicly available at: \texttt{\href{https://huggingface.co/inpars-plus}{huggingface.co/inpars-plus}}.
\end{itemize}

\section{Background and Preliminaries}


In this section, we summarise the key concepts behind the QGen pipelines and the methods used in the InPars Toolkit. We also briefly introduce the main contributions of the InPars Toolkit and highlight the key limitations that motivate this work. Finally, we give a summary of the relevant literature we used in our extensions.

\subsection{QGen Pipelines}
All synthetic query generation (QGen) pipelines considered in this work, as well as in \cite{abonizio2023inpars}, have a similar structure. First, a generator LLM is presented with a prompt containing a target document and a set of relevant query-document pairs. The generator LLM is then invoked to generate a relevant query for this document. This process is repeated for each document in the dataset. The generated queries are subsequently filtered based on a scoring mechanism, and the high-quality queries are used to fine-tune a reranker model. At the end of the pipeline, the fine-tuned reranker model is evaluated on some test data and various retrieval metrics are reported.

Depending on which LLMs are used for query generation, which prompting strategies are employed as well as the filtering mechanism, the InPars Toolkit \cite{abonizio2023inpars} considers the following three QGen pipelines: InPars-V1 \cite{bonifacio2022inpars}, InPars-V2\cite{jeronymo2023inpars}, and Promptagator\cite{dai2022promptagator}.

\subsection{InPars}\label{sec:inpars}
The initial InPars paper, later referred to as InPars-V1, proposed by Bonifacio et al. 2022, introduced an effective method for leveraging large LLMs in retrieval by utilizing them as generators of synthetic data. The authors identified that using LLMs directly during the retrieval stage is infeasible; thus, they proposed shifting the computational cost to the synthetic data generation process for training. 

The InPars method creates a prompt by concatenating a static prefix $t$ with a document from the target domain $d$. InPars considers two different (fixed) prompt templates: a \textit{vanilla} template and a \textit{guided by bad question (GBQ)} template. The \textit{vanilla} template consists of a fixed set of 3 pairs of queries and their relevant documents, sampled from the MS MARCO \cite{bajaj2016ms} dataset. The \textit{GBQ} prompt extends this format by posing the original query to be a bad question, in contrast with a good question manually constructed by the authors. Feeding this prefix-target document pair $t||q$ to the LLM is then expected to output a novel query $q^*$ likely to be relevant to the target document. Thousands of these positive examples are generated for the target domain, and are later used to fine-tune a monoT5 reranker model \cite{nogueira2020document}.

However, the authors identified that using this full set of generations for training does not yield optimal results. This is likely owed to the poor quality of a large portion of the generated examples. Therefore, the authors propose to apply a scoring mechanism to select the top $K$ of generations based on the following score:

\begin{equation}
    p_q = \cfrac{1}{|q|}\sum_{i=1}^{|q|} \log p(q_i|t, d, q_{<i})
\end{equation}

where $p(q_i|t, d, q_{<i})$ represents the probability of generating token $i$ from generated query $q$ assigned by the generator LLM, GPT-3 in the case of InPars.

As a folloup work, InPars-v2 \cite{jeronymo2023inpars} extends the original InPars by replacing the scoring metric with a relevance score provided by a monoT5-3B reranker model. Additionally, the authors switch the generator LLM to GPT-J \cite{wang2021gpt}.

In both works, the authors prompt the generator model 100\,000 times for synthetic queries but only keep the top 10\,000 highest scoring instances. This means that 90\% of the generated queries are discarded. It should be noted that the authors provide no substantiation or ablation for this setting.


\subsection{Promptagator}
The Promptagator method, proposed by Dai et al. 2022, operates similarly to the InPars method. Promptagator prompts the 137B-parameter model FLAN \cite{wei2021finetuned} with eight query-document examples, followed by a target document, to generate a relevant query for the target document. Unlike InPars, which uses a fixed prompt template, Promptagator employs a dataset-specific prompt template tailored to the target dataset's retrieval task. 

Subsequently, high-quality generations are filtered from the batch of outputs using a process called consistency filtering \cite{alberti2019synthetic, lewis2021paq}. Consistency filtering ensures that a query is answered by the passage from which it was generated. To this end, a retrieval model is invoked and the generated query is only accepted if the target document appears in the top-K retrieved documents.

Abonizio et al. (2023) highlight several limitations in these works. Firstly, neither InPars nor Promptagator are fully reproducible, partly due to the use of models that are not publicly available. Additionally, the InPars models are restricted to specific hardware (TPUs), making them difficult to adapt to other models. Lastly, the Promptagator method lacks a public code base entirely. To address these limitations, Abonizio et al. introduce the InPars Toolkit, an open-source end-to-end and fully reproducible pipeline for synthetic data generation with GPU support (via PyTorch~\cite{paszke2019pytorch}).

\subsection{Contrastive Preference optimization}
To address the performance gap between moderate-sized LLMs (7–13 billion parameters) and state-of-the-art large LLMs, such as GPT-4 \cite{openai2023gpt}, in the task of machine translation, Xu et al. proposed Contrastive Preference optimization (CPO) \cite{xu2024contrastive}.

The goal of CPO is to overcome two key shortcomings of supervised fine-tuning. First, the performance of supervised fine-tuning is limited by the quality of gold-standard human-annotated reference translations. Second, supervised fine-tuning cannot reject errors present in the provided gold-standard translations.

CPO addresses these limitations by introducing a training schema that utilises triplets comprising reference translations, translations from a teacher model (e.g., GPT-4), and those from the student model. Reference-free evaluation models are employed to score the translations, and these scores are subsequently used to guide the student model toward generating preferred translations.

In practice, contrastive preference optimization forms an approximation of Direct Policy Optimization by instead minimizing its upper bound. 

Given a set of source sentences $\mathcal{X}$, alongside preferred targets $\mathcal{Y}_w$ and less preferred ones $\mathcal{Y}_l$, we can construct a triplet dataset of input, preferred output, and dispreferred output, which we denote as \(\mathcal{D} = \{x^{(i)}, y^{(i)}_w, y^{(i)}_l\}_{i=1}^N\). Then, the preference optimization loss term is defined as: 

\begin{align*}
L(\pi_\theta; U) = -\mathbb{E}_{(x, y_w, y_l) \sim \mathcal{D}} & [ \log \sigma ( \beta \log \pi_\theta(y_w|x) \\  &- \beta \log \pi_\theta(y_l|x) )]
\end{align*}

where U is a uniform prior policy which replaces $\pi_\text{ref}$ in classical DPO. This relaxation allows for only storing and requiring computations for the target policy model -- effectively reducing computational and memory requirements in half. More importantly, because the preferred and dispreferred targets are determined by an objective (reference-free) metric, they are potentially able to approximate better the optimal policy $\pi^*$. In the context of machine translation, this trains the model to avoid generating adequate but imperfect outputs.     

Furthermore, the authors \cite{xu2024contrastive} incorporate a behaviour-cloning (BC) regularizer such that the distribution of the target model's generations does not deviate too far from the distribution of the teacher model's generations, which is defined as:

\[
\mathcal{L}_{NLL} = \mathbb{E}_{(x, y_w) \sim D} \left[ \log \pi_\theta(y_w|x) \right]
\]

Thus, the final CPO objective is defined as:

\[
\min_\theta
\underbrace{L(\pi_\theta, U)}_{\mathcal{L}_\text{prefer}} - \underbrace{\mathbb{E}_{(x, y_w) \sim D} \left[ \log \pi_\theta(y_w|x) \right]}_{\mathcal{L}_\text{NLL}}
\]

\subsection{DSPy}\label{sec:dspy}
The rise of promptable LLMs has inevitably led the research community to explore techniques for effectively prompting these models. Consequently, LLM pipelines often rely on hard-coded prompt templates, of which the InPars Toolkit is an example. To address this limitation, Khattab et al. introduced DSPy \cite{khattab2024dspy}, a programming model that offers a more systematic approach to optimizing LLM pipelines. DSPy brings the construction and optimization of LLM pipelines closer to traditional programming, where a compiler automatically constructs prompts and invocation strategies by optimizing the weights of general-purpose layers following a program.

\subsection{Research Gap}
In light of the previously discussed studies, it becomes apparent that synthetic data generation pipelines, such as InPars and Promptagator, have proven to be effective strategies for improving Neural Information Retrieval models. Previous studies have focused on enhancing different components of these pipelines. For instance, InPars-V2 has improved the scoring mechanism used for synthetic data filtering, InPars-light has enhanced overall efficiency by exclusively using more lightweight, open-source models, and InRanker has addressed the reranking bottleneck by distilling knowledge from a large MonoT5-3B model specialized in the ranking task into smaller counterparts while utilizing synthetic data from InPars.

However, to the best of our knowledge, no work has addressed the issue of reducing noise in the generated queries. In the case of InPars, 90\% of the generated queries are filtered out and not used in subsequent stages of the pipeline. Since query generation comprises one of the most computationally expensive stages, this study aims to mitigate this inefficiency. We explore replacing the static prompt templates used in the InPars Toolkit pipeline with DSPy prompt optimization, with the aim of improving the query generation process. Additionally, we investigate applying a knowledge distillation strategy, somewhat similar to InRanker, to the generator LLM using a CPO fine-tuning procedure. 

\section{Methodology}

In this section, we begin by discussing our approach to ensuring reproducibility of the claims of the original authors, then we outline our methodology for extending the work of the original authors.

\subsection{Reproducibility}
As specified in Section 1, we start by assessing the reproducibility of InPars Toolkit. To this end, we aim to verify the validity of the three main claims we have identified: (1) the InPars Toolkit is an end-to-end reproducible pipeline; (2) it provides a comprehensive guide for reproducing InPars-V1, InPars-V2, and partially Promptagator; and (3) the toolkit has plug-and-play functionality.
Given the substantial computational resources required to fully reproduce all experiments on the 18 BEIR benchmark datasets as proposed by Abonizio et al.\cite{abonizio2023inpars}—approximately 2000 GPU hours—we opted for a more efficient approach. Fortunately, the authors have made a significant portion of the synthetic data and fine-tuned models publicly available\footnote{\href{https://huggingface.co/inpars}{InPars on HuggingFace} \\ \href{https://huggingface.co/models?search=zeta-alpha-ai+monot5}{InPars fine-tuned monoT5 models} \\ \href{https://github.com/zetaalphavector/InPars?tab=readme-ov-file\#Resources}{Listed resource on the GitHub repository}\\ }. To reduce the energy footprint of this reproducibility study, we propose conducting three different types of reproduction experiments on the smallest dataset in the BEIR benchmark—SciFact \cite{scifact}. 

\begin{figure}[hbpt]
    \centering
    \includegraphics[width=0.9\linewidth]{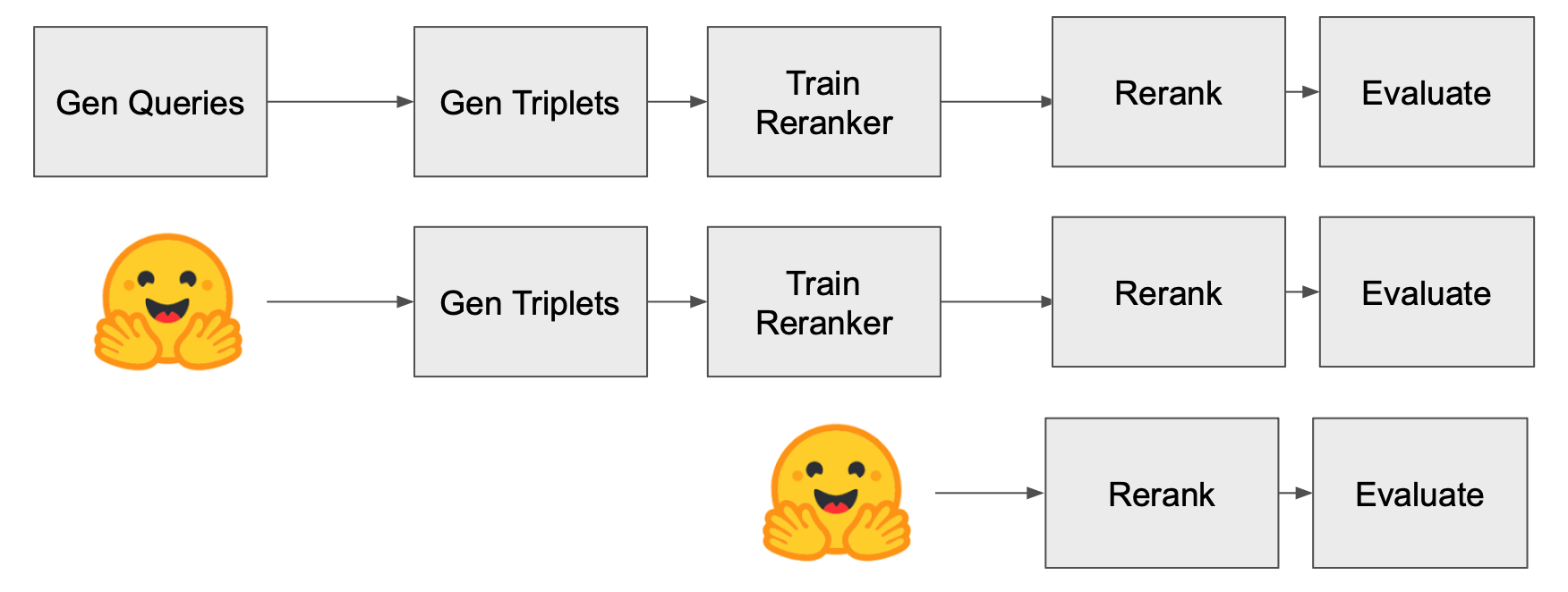}
    \caption{Diagram of the three different reproducibility experiments.}\label{fig:repro}
\end{figure}

Our first experiment will be a full end-to-end reproduction of the InPars Toolkit methodology for reproducing InPars-V1, InPars-V2, and partially Promptagator, thereby assessing the validity of Claims 1 and 2. Secondly, we skip the expensive data generation stage by downloading the synthetic data made publicly available by the authors. By doing so, we assess the validity of the published resources while providing additional evidence of the InPars Toolkit's reproducibility. Finally, we will utilise the pre-trained ranker models published by the authors for reranking.

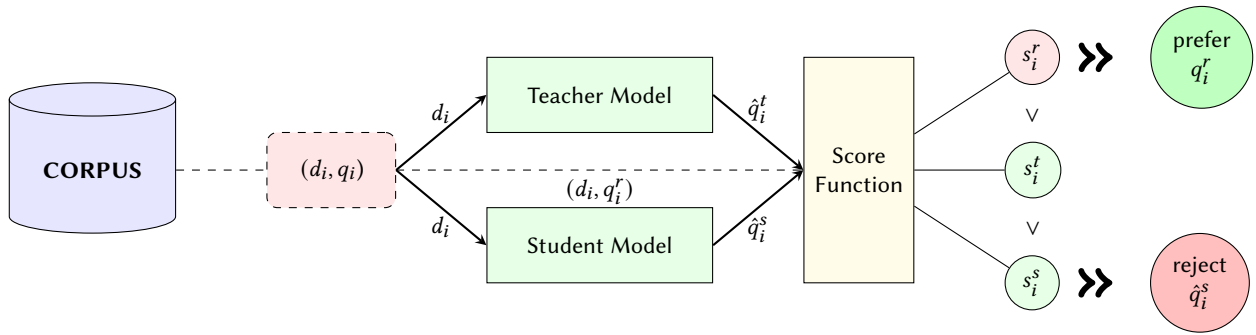
\begin{figure*}[htbp]
\centering

\begin{tikzpicture}[
    font=\sffamily,
    node distance=1.96cm and 1.2cm,
    every node/.style={align=center},
    doc/.style={draw, rounded corners, fill=red!10, text width=1.5cm, minimum height=1cm, align=center},
    model/.style={draw, rounded corners, fill=blue!10, text width=2.5cm, align=center},
    scoring/.style={draw, dashed, rounded corners, fill=orange!10, text width=2cm, align=center},
    output/.style={circle, draw, fill=green!10, minimum size=0.5cm, align=center}]

\tikzstyle{database} = [cylinder, draw, shape border rotate=90, aspect=0.2, 
    minimum height=2cm, minimum width=1cm, text width=2cm, align=center, fill=blue!10]
\tikzstyle{model} = [rectangle, draw, minimum width=3cm, minimum height=1cm, text centered, fill=green!10]

\tikzstyle{arrow} = [thick,->,>=stealth]

\node[database] (db) {\textbf{CORPUS}};
\node[doc, right=of db, dashed, draw, minimum width=0.5cm] (docquery) {$(d_i, q_i$)};
\node[model, right=of docquery, yshift=1cm] (teacher) {Teacher Model};
\node[model, right=of docquery, yshift=-1cm] (student) {Student Model};
\node[model, right=of teacher, yshift=-1cm, fill=yellow!10, minimum height=3cm,  minimum width=0.5cm, text width=1.25cm] (scoring) {Score Function};

\draw[dashed] (db) -- (docquery);
\draw[arrow] (docquery.east) -- (teacher.west) node[midway, above] {$d_i$};
\draw[arrow] (docquery.east) -- (student.west) node[midway, below] {$d_i$};
\draw[arrow] (teacher.east) -- (scoring.west) node[midway, above] {$\hat{q}^t_i$};
\draw[arrow] (student.east) -- (scoring.west) node[midway, below] {$\hat{q}^s_i$};
\draw[dashed] (docquery.east) to node[midway, below] {$(d_i, q^{r}_i)$} (scoring.west);
\node[output, right=of scoring, yshift=1.5cm, fill=red!10] (output0) {$s^r_i$};
\node[output, right=of scoring, yshift=0cm] (output1) {$s^t_i$};
\node[output, right=of scoring, yshift=-1.5cm] (output2) {$s^s_i$};

\node[output, right=of output0, fill=green!25] (prefer) {prefer \\ $q^r_i$};
\node[output, right=of output2, fill=red!25] (reject) {reject \\ $\hat{q}^s_i$};
\draw[->>, double, double distance = 2pt, shorten >= 5mm, shorten <= 6.6 mm] (output0) -- (prefer);
\draw[->>, double, double distance = 2pt, shorten >= 5mm, shorten <= 6.6 mm] (output2) -- (reject);

\draw (scoring) -- (output0);
\draw (scoring) -- (output1);
\draw (scoring) -- (output2);

\node[below=of output0, yshift=1.75cm] (gt0) {\textbf{$\vee$}};
\node[below=of output1, yshift=1.75cm] (gt1) {\textbf{$\vee$}};

\end{tikzpicture}

\caption{Data generation for CPO-based fine-tuning. A relevant document-query pair is sampled from the dataset. Both teacher and student models generate queries for the given document, yielding three queries: $q^r_i$, $\hat{q}^t_i$, and $\hat{q}^s_i$. These queries are scored for similarity against the target document, with the highest-scoring query designated as preferred and the lowest-scoring as rejected. This procedure is applied across all relevant query-document pairs in the corpus.}
\label{fig:cpo_gen}
\end{figure*}

A minor issue with the published resources is that not all of them were found, despite the authors claiming to have published them. Specifically, we were unable to locate the synthetic data generated using the Promptagator prompt templates specified in the InPars Toolkit, nor did we find the pre-trained reranker models trained on InPars-v1 generations. Consequently, these experiments were not conducted. 

To assess the validity of claim 3, regarding plug-and-play functionality, we additionally perform a full end-to-end run of the QGen pipeline using a newer LLM, namely Llama 3.1 \cite{llama3} for query generation. Since the original authors experiment with the GPT-J-6B model, we attempt to match this with the 8B parameters version of LLama 3.1. We expect this model to seamlessly integrate into the InPars Toolkit pipeline and potentially improve downstream performance.

To further reduce the computational workload, all experiments are run using half precision (FP16). Furthermore, we use the default parameters specified by the authors in the reproducibility guide and the published code. 

\subsection{Extending InPars-toolkit}

\subsubsection{Fine-tuning the Generator}
We hypothesize that targetting the generator model for fine-tuning individually will result in better retrieval performance, and will alleviate the issue of wasted compute. Furthermore, given high enough quality, we may even omit filtering altogether. 

Since our main goal is to reduce the total cost of the QGen pipeline, we experiment with CPO to obtain a model that generalizes to the task of query generation across different target domains. This approach aims to minimize wasteful computation and hopefully eliminates the need to re-train the model for each target dataset.


Following Xu et al. 2024, we adapt the triplet data generation pipeline by proposing the following modifications for synthetic query generation (also pictured in Fig. \ref{fig:cpo_gen}).

\begin{enumerate}
    \item Sample N pairs of \textit{relevant} query-document pairs, preferably with all available documents in the corpus. 
    \item Using two models -- one teacher and one student, generate predicted queries for each document. Here we also employ targeted prompt templates, as described in §\ref{sec:targetedtemplates}.
    \item Compute a relevance score between the three queries (one from student, one from teacher, and one as the reference) and document, as described in §\ref{sec:query_eval}.
    \item Filter out irrelevant data samples for which all scores lie within preset margins $L < s^i_{s/t/r} < H$. This aims to reduce degenerate cases where the model outputs copies of the document or irrelevant queries. We use $L=0.3$ and $H=0.7$. 
    \item Select the query corresponding to the highest relevant score as the ``preferred'' example to optimize for, and the query corresponding to the lowest score as the ``dispreferred'' example.
    \item Train using CPO, following Xu et al~\cite{xu2024contrastive}.
\end{enumerate}

The trained generator model can then be used in zero-shot fashion in order to produce higher quality queries, enhancing downstream performance. 

For our study, we employ Llama 3.1 8B Instruct\footnote{\url{https://huggingface.co/neuralmagic/Meta-Llama-3.1-8B-Instruct-FP8}}, as well as Llama 3.1 Nemotron 70B Instruct\footnote{\url{https://huggingface.co/neuralmagic/Llama-3.1-Nemotron-70B-Instruct-HF-FP8-dynamic}} \cite{llama3} for our student and teacher models, respectively.  We hypothesize that the knowledge distillation effect of preference-based optimization will be strongest on models of similar architecture. We fine-tune a single student model on 100,000 samples from the MS MARCO passage dataset and aim to evaluate its QGen performance on a subset of the BEIR benchmark.

\subsubsection{Targeted Prompt Templates} \label{sec:targetedtemplates}
Furthermore, we incorporate targeted prompt templates inspired by Promptagator and enhance them by providing K in-distribution examples, following the InPars method. We also refer to this prompting strategy as \textsc{InPars+} prompts.

Due to time and budget constraints, we opt to not use the CoT methodology outlined in §\ref{sec:dspy} to build the prompts for generator fine-tuning. However, any observed benefits of the DSPy method can potentially carry over well, as the two approaches carry theoretically multiplicative benefits.   

\subsubsection{Query Evaluation}\label{sec:query_eval}

Preference optimization in general requires a scoring mechanism that determines the weight of the preferred or dispreferred option. In the context of information retrieval, we employ a scoring function combining Siamese networks for retrieval with normalized BM25 scores. The latter serves to alleviate some of the issues of using bi-encoders for similarity scoring. 
The text encoder score $s_{enc} \in [0,1]$ is defined as

\[
 s_{enc} (\text{doc, query}) = 1 + \frac{h_{\text{doc}} \cdot h_{\text{query}}}{2 ||h_{\text{doc}}|| \cdot ||h_{\text{query}}||}
\]

where $s_{enc}$ is the re-scaled cosine similarity between the embedding vectors produced by the Siamese networks for query and document. In our experiments, we employ Sentence Transformer \cite{reimers-gurevych-2019-sentence} encoder where the output of the last Transformer layer is aggregated with a mean pooling layer. Document embeddings are precomputed and stored in an embedding table to accelerate retrieval.

In a similar fashion, we precompute the IDF index for BM25, and compute BM25 scores across the batch of queries on-the-fly. Because BM-25 scores are inherently not normalized, we take the softmax across the entire corpus and only get the score corresponding to the target query-document pair. The final score $s$ is calculated as:
\[
s(\text{doc, query}) = 0.5 \cdot s_{enc}(\text{doc, query}) + 0.5 \cdot s_{\text{BM25}}(\text{doc, query})
\]

\begin{figure}
    \centering
    \begin{subfigure}{\linewidth}
    \includegraphics[width=\linewidth]{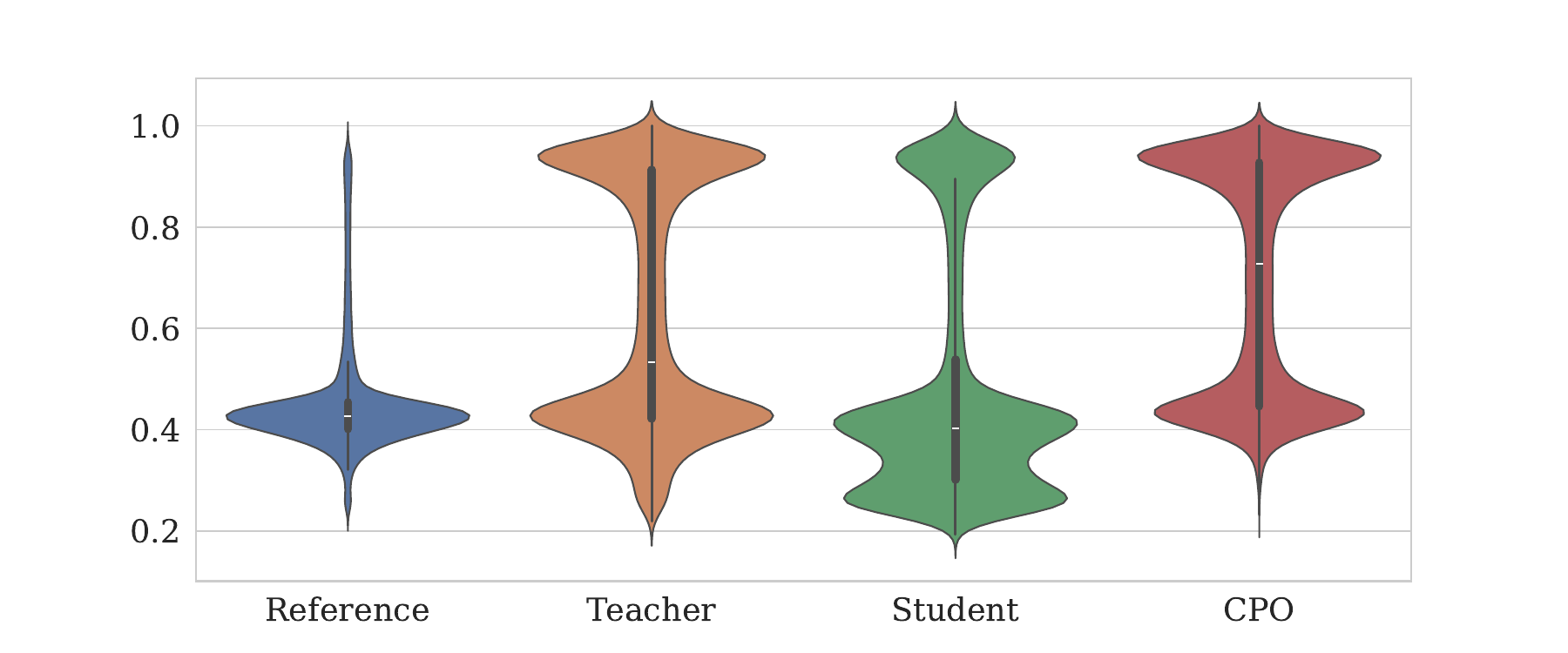}
    \caption{MS MARCO}
    \end{subfigure}
    \begin{subfigure}{\linewidth}
        \includegraphics[width=\linewidth]{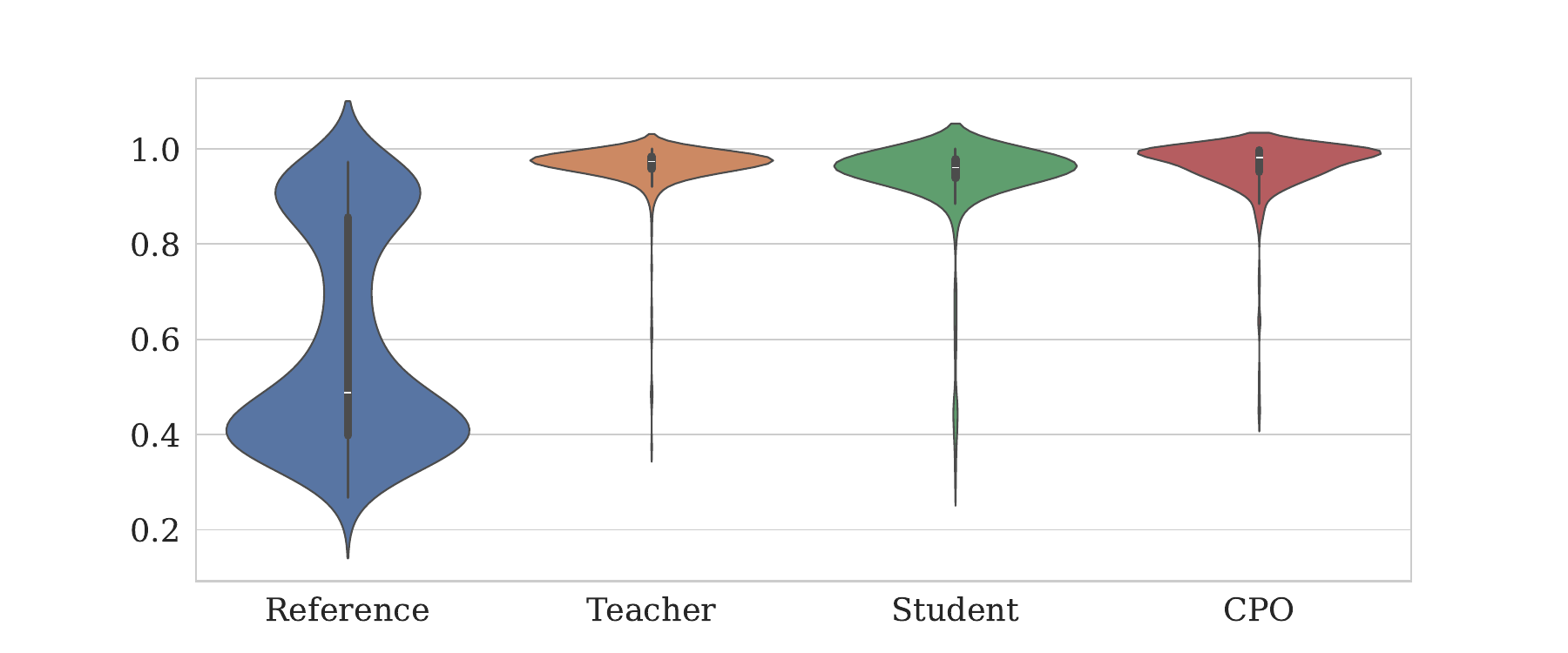}
        \caption{SCIFACT}
    \end{subfigure}
    \caption{Comparison of similarity score distribution with query evaluation applied on two sample datasets using Llama 3.1 70B as the teacher model and Llama 3.1 8B as the student model. In a very noisy dataset such as MS MARCO, both teacher and student are able to generate higher- or similar-quality queries compared to the reference. Post-CPO, the student is able to generate queries with a higher mode.}
    \label{fig:query_eval}
\end{figure}

\subsubsection{Pipeline Enhancements}
In addition, we implement many other enhancements to better utilize the available resources on the system running InPars.

\begin{itemize}
    \item Higher CPU utilization during prompt generation using the Prompt builder class. On systems with high number of logical processing cores, this substantially reduces the time required for generating prompts.
    \item Additional caching and intermediate result backup, allowing easy recovery from program crashes.
    \item Enhanced inference using vLLM~\cite{kwon2023efficient}, enabling significant speed-ups and better memory utilization for multi-billion parameter models. In particular, this also allows scaling up the number of synthetic queries within a similar budget which can in turn yield improvements in downstream retrieval performance.
\end{itemize}

\subsubsection{DSPy for Enhanced Prompt Generation}
To further improve the quality of generated queries, we also study incorporating the DSPy framework \cite{khattab2024dspy}. Traditional synthetic data generation approaches, such as InPars, rely on static, handcrafted prompt templates that do not scale well to unseen datasets. To address this issue, we leverage the LLM’s own capabilities to determine what constitutes a relevant query. We utilize Chain of Thought (CoT) reasoning to guide the LLM in breaking down the document’s content into a series of logical steps before formulating the final query. Prior work has shown that CoT reasoning can enhance performance by encouraging more structured, contextually appropriate outputs~\cite{wei2022chain}. 

Following this approach, we treat the LLM as an agent by providing it a set of instructions to aid its adaptive query generation, ensuring the model reasons about what is relevant to the dataset.

\begin{figure}[!hbpt]
    \centering
    \includegraphics[width=0.82\linewidth]{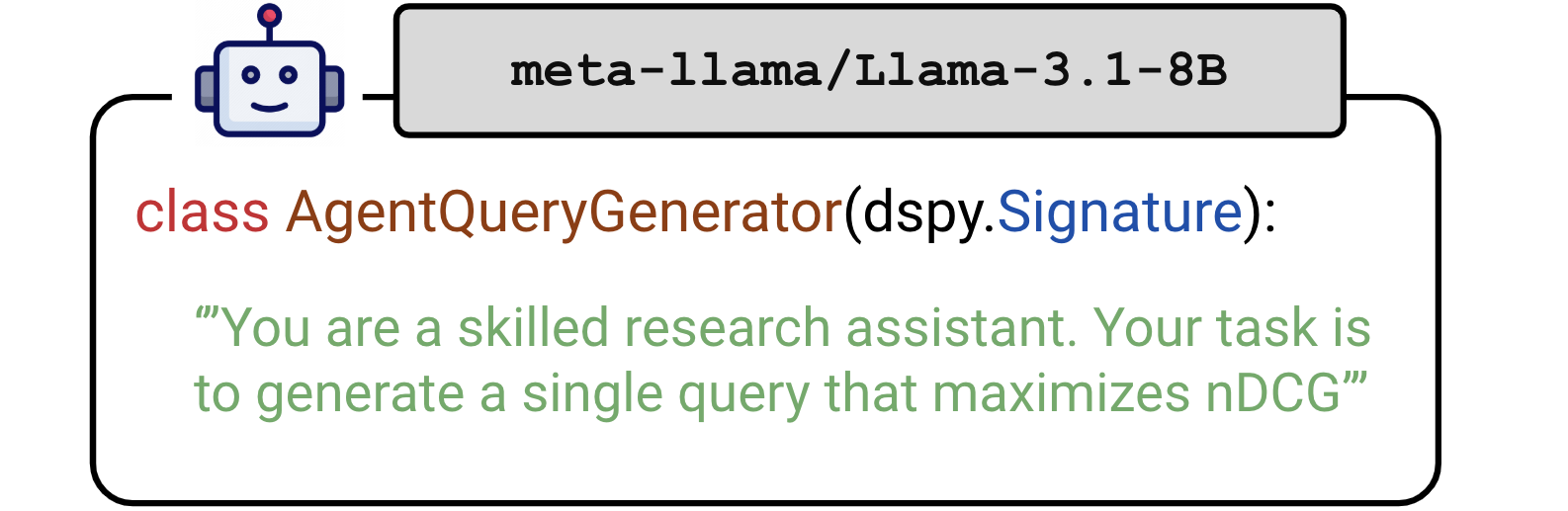}
    \caption{Module definition for AgentQueryGenerator, a DSPy signature that prompts a meta-llama/Llama-3.1-8B model to behave as a skilled research assistant.}
\label{fig:agent}
\end{figure}

Figure \ref{fig:agent} illustrates the signature that configures the LLM as an agent. Rather than relying on a fixed prompt template, DSPy uses this signature to dynamically create a prompt, embedding the agent’s instructions directly into the model’s reasoning steps. 

While CoT allows the model to analyze the document systematically before generating queries, this came at the cost of additional computations. 

Furthermore, the model demonstrated a tendency to deviate from its primary task -- instead producing outputs which resemble a continuation or extrapolation of the document and prompt. To alleviate this, we introduce stopping words which ensures adherence to the task while preserving the integrity of the CoT approach.

\subsubsection{Impact of Filtering}
Finally, we hypothesize that an improved query generation process should enable reduced filtering, leading to greater computational efficiency. Thus, we analyze the effect of filtering using Zero-Shot and CoT strategies by conducting a full end-to-end experiment, reducing the number of generated queries from 100k to 10k, 25k, and 50k, thereby simulating scenarios where significantly fewer or no generations are removed by filtering.

\section{Results}
In the following sections, we describe the reproduction of the original InPars results and discuss the effects of our extensions.

\subsection{Reproduction Experiments}
The results of the reproductions are listed in Table \ref{tab:comparison}. We observe slight delta between our reproduction from scratch (experiment 1) and results reported by the authors, where our results are consistently lower for the target dataset. 

With the synthetic query dataset provided by the authors (experiment 2), our results are approaching the reported ones, which might indicate some inconsistencies stemming from the dataset generation procedure. This assumption is validated when we also utilize their fine-tuned reranker (experiment 3), where both results match the reported results for the InPars-V2 pipeline.

Furthermore, using a bigger and more selectively trained model \hspace{0.1cm}~\cite{llama3}, yields some marginal gains in downstream performance, as seen in the columns listing the downstream results of using synthetic queries generated by GPT-J and LLaMA. 

In summary, our results largely match the original authors', despite some minor inconsistencies. This validates the identified claims that the pipeline is end-to-end reproducible and does incorporate plug-and-play functionality for different generator and reranker models with minimal modifications. 

\begin{table}[htbp]
    \centering
    \caption{Results of the reproduction experiments for the SciFact dataset. The $\dagger$ symbol indicates experiments using Promptagator templates. The number (1) in the second and third columns indicate we have conducted experiment 1: full pipeline.} \label{tab:comparison}
    \small
    \begin{tabular}{lccccc}
        \toprule
        & \textbf{InPars} & \textbf{GPT-J (1)} & \textbf{LLaMA (1)} & \textbf{Exp.(2)} & \textbf{Exp.(3)} \\
        \midrule
        \textbf{BM25}             & 0.678 & 0.679 & 0.679 & 0.679 & 0.679 \\
        \midrule
        \textbf{InPars-V1}        & 0.774 & 0.758 & 0.759 & 0.770 & --     \\
        \textbf{InPars-V2}        & 0.774 & 0.752 & 0.759 & 0.770 & 0.770 \\
        \textbf{InPars-V1$^\dagger$} & 0.790 & 0.766 & 0.778 & --     & --     \\
        \textbf{InPars-V2$^\dagger$} & 0.790 & 0.769 & 0.786 & --     & 0.782 \\
        \midrule
          \textbf{Average} & \textbf{0.782} & \textbf{0.761} & \textbf{0.771} & \textbf{0.770}     & \textbf{0.776} \\
         \bottomrule
    \end{tabular}
\end{table}

\subsection{Extensions and Ablations}

\begin{table*}[!htbp]
    \centering
    \caption{Main experimental results. We generate 10,000 synthetic queries using the model specified in the first column with InPars-V2 filtering (SC, Sec.\ref{sec:inpars}) and fine-tune a monoT5 reranker for one epoch. For each test query, we sample the top-1000 documents via BM25 and rerank using the fine-tuned monoT5. (1) indicates no reranking baseline. For (5): the first row shows our fine-tuned LLama3.1 8B model using CPO on MSMARCO with original InPars prompting; the second row employs InPars+ prompts (Sec\ref{sec:targetedtemplates}); the third row uses CoT prompting.}
    \label{tab:main_results}
    \begin{tabular}{clcccccc}
        \toprule
          & \textbf{Method}& \textbf{Filter} & \textbf{\textsc{Cov}} & \textbf{\textsc{NFC}}  & \textbf{\textsc{Arg}} & \textbf{\textsc{Scifact}}&\textbf{\textsc{Avg}}\\
        \midrule
         (1) & BM25 & N/A & 0.595& 0.322&  0.397& 0.679&0.498\\
        \midrule
         (2) & GPT-J 6B & SC   & 0.824& 0.373& 0.105& \underline{0.770}  &0.518\\
         \midrule
         (3.1) & Llama 3.1 8B  & SC & {0.845}& \underline{0.380}& 0.126& 0.759& {0.531}\\
         (3.2) & Llama 3.1 70B  & SC & 0.708& 0.378& 0.113& 0.754& 0.488\\
         \midrule
         (4) & Llama 3.1 8B + CoT & SC & \underline{0.856}& \textbf{0.390}& 0.368& \textbf{0.786}&{0.600}\\
         \midrule
         (5) & CPO @ MSMARCO & SC & 0.778 & 0.372 & 0.253 & 0.746 & 0.490\\

          & \hspace{0.05in} w/ IP+ prompts & SC & 0.769 & 0.368 & \bfseries 0.609 & 0.749 & \bfseries 0.622\\
          & \hspace{0.05in} w/ DSPy CoT & SC & \bfseries 0.867 & 0.371 & \underline{0.417} & 0.761 &  \underline{0.604}\\

         \bottomrule
    \end{tabular}
\end{table*}

\subsubsection{Impact of Filtering}
To evaluate the effectiveness of DSPy prompts and the impact of filtering, we propose an experiment simulating low-resource settings. All fine-tuned reranker models in our experiments were trained on a subset of 10,000 synthetic queries. In the original experiment proposed by the authors of InPars, 100k query generations were filtered to produce a smaller subset of 10k queries. To investigate the effects of filtering and assess the potential efficiency gains of using DSPy for prompt construction, we propose reducing the initial pool of 100k generations to smaller random samples of 10k (no filtering), 25k, and 50k queries, thereby adjusting the filter ratio from 90\% to 0\%, 60\%, and 80\%, respectively.

\begin{figure}[!hbpt]
    \centering
    \includegraphics[width=\linewidth]{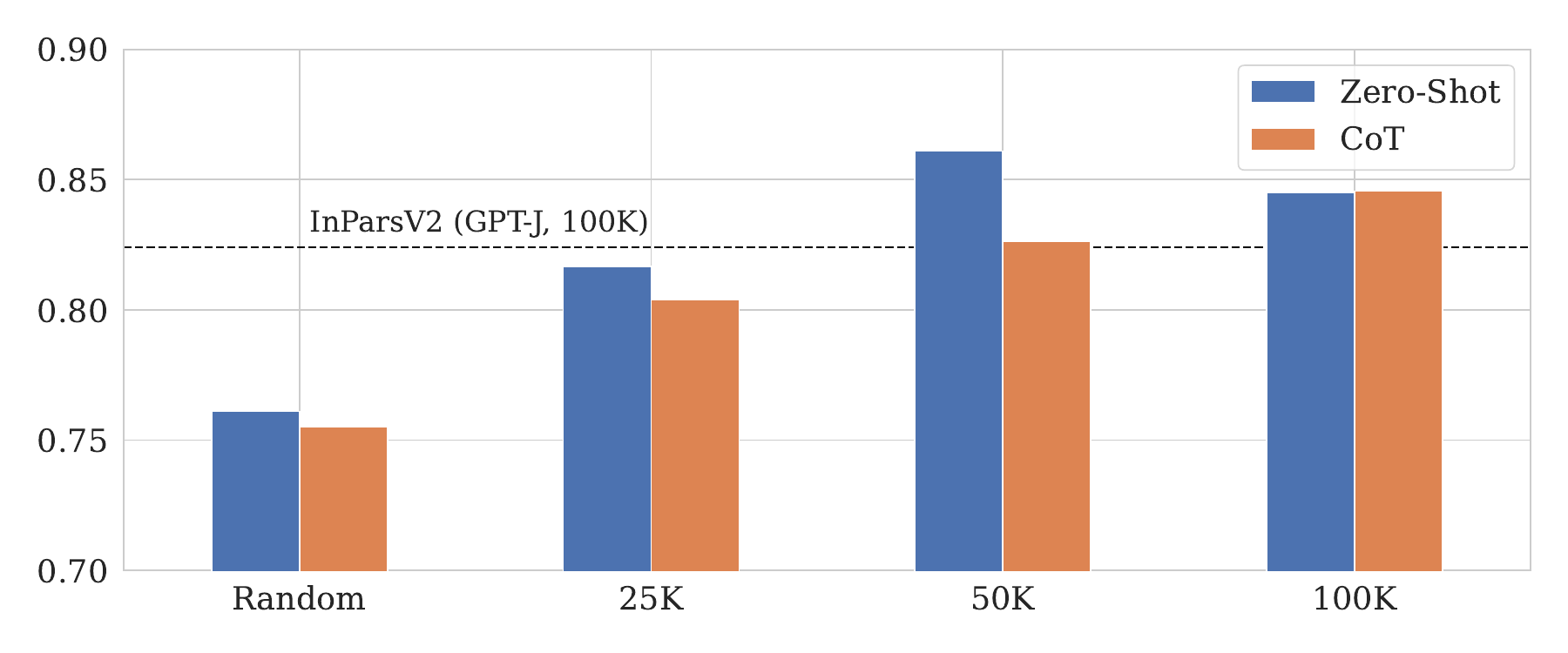}
    \caption{Impact of filtering for zero-shot generation and CoT-derived generation using monoT5 as reranker (K=1000). The dotted line represents the baseline performance
    using the top 10,000 queries from GPT-J, while the bars depict
    results for queries generated by Llama 3.1 8B across various subsets.}
    \label{fig:filter_impact}
\end{figure}

In this experiment, (Fig. \ref{fig:filter_impact}), we adopt the InPars-V2 strategy, where the top 10k queries are selected based on scores generated by a MonoT5-3B reranker model. We experiment with the Trec-Covid dataset.

We find that filtering from larger subsets generally enhances the quality of the trained reranker. This improvement likely stems from the larger pool of instances, offering more opportunities to select high-quality queries. Notably, the same results indicate that CoT-produced prompts lead to degraded performance. Furthermore, we observe an unexplained bump in performance when filtering from the 50k subset, which persists across different seeds. Overall, the results demonstrate that our proposed CoT prompting strategy, combined with the 8B model, achieves performance comparable to the original InPars-V2 results while using only a quarter of the queries. These results suggest that CoT can enhance efficiency and, under equivalent resource conditions, even improve downstream performance. 

It is important to note that comparing experiments conducted with Llama to those using GPT-J is not a direct one-to-one comparison and may influence the results. 

\subsubsection{Results in CPO}
We observe that a trained student was able to learn to generate generally higher scoring queries (Fig. \ref{fig:query_eval}), pushing the score distribution upwards. However, as we observe in our experimental results, bigger models do not necessarily produce better outputs and this is also reflected on the CPO learner.

The model trained on the MS MARCO subset seems to perform reasonably well as a generalist generator. Furthermore, we observe that filtering for these models resulted in generally worse performance for the reranker. As above, we hypothesize that we are likely removing useful training signal. This seems particularly pronounced for datasets such as Arguana, as there is only a limited set of training data available.

\subsubsection{DSPy results}
Similarly, the use of dynamic prompt optimization combined with Chain-of-Thought (CoT) consistently improves performance across datasets. When compared to the baseline, it is clear that the base Llama model, when used with DSPy and InPars-V2 score-based filtering, achieves the highest average score. Notably, for the Arguana dataset, although BM25 still outperforms the other baselines, the CoT approach significantly enhances performance, raising the nDCG score from 0.126 to 0.368, which is an improvement of 27 nDCG@10 points.

\subsubsection{Impact of Reranker model}
To further investigate the effectiveness of our approaches, we target ranking fewer documents ($K=100$) and using a much smaller reranker model, in our case \textsc{miniLM}\footnote{https://huggingface.co/cross-encoder/ms-marco-MiniLM-L-6-v2}, which has been shown to work comparatively well for how lightweight it is\cite{boytsov2023inpars}. We observe a similar trend (Fig.~\ref{tab:scifact_minilm_top100}) as with reranking $K=1000$ documents and monoT5 -- filtering can be beneficial for some domains and harmful for others. Furthermore, bigger and theoretically more capable models do not necessarily result in improved downstream performance.

\begin{table}[h!bp]
  \centering
  \caption{nDCG@10 results for \textsc{Scifact} and \textsc{TREC-Covid} using MiniLM and reranking the top-100 documents. NF stands for no filter, and SC stands for ``scores'' filtering adapted from InPars-V2~\cite{jeronymo2023inpars}}
  \label{tab:scifact_minilm_top100}
  \begin{tabular}{lccc}
    \toprule
     Generator & Filter & \textsc{Scifact}& \textsc{Covid}\\
    \midrule
    Llama 3.1 8B & NF  & 0.707 & 0.775  \\
     & SC & 0.719 & 0.733 \\
    Llama 3.1 70B &NF & 0.700 & 0.763 \\
     & SC & 0.745 & 0.728 \\
    \bottomrule
    \end{tabular}
\end{table}

\vspace{-0.5cm}

\section{Discussion}

\subsection{Query Evaluation}
We provide a quantitative demonstration of the query evaluation framework in Figure \ref{fig:query_eval} and contextualized samples in Table \ref{tab:query_eval}. Although the fine-tuned model is able to learn to generate better quality queries, it can occasionally repeat the entire document, which is not useful for training but results in a high query evaluation score (See Sec~\ref{sec:query_eval}). To mitigate this, we introduced a maximum threshold for the query evaluation score to avoid training on such false-positives. For future work, we aim to explore more sophisticated approaches for query evaluation.


\subsection{Contrastive Preference Optimization}
One key assumption we made throughout this paper was that using better generator models will translate into higher quality of downstream performance for our reranker using synthetic data. However, this does not hold true in general, as seen in rows (3.1) and (3.2) in Table \ref{tab:main_results}.  


We determine that further research is essential to determine the potential impact of optimizing synthetic data generator models for information retrieval. In particular, we plan to investigate better ways to generate training datasets to encourage more appropriate query generation. 

\subsection{CoT Prompts}
As shown in Table \ref{tab:main_results}, incorporating a guided process improves query quality; however, this comes with trade-offs and requires careful consideration of certain aspects.

While we anticipated that applying CoT would improve query quality and reduce the number of instances required for training, this was not consistently observed, particularly for the TREC-Covid dataset in figure \ref{fig:filter_impact}. This could be due to several factors:

\begin{itemize}
    \item The TREC-Covid dataset represents a corpus of 171k documents. Training on a subset of the highest scoring documents seems to decrease the performance of the reranker. Training on higher number of samples, instead of filtering, seems to provide better training signal translated into better downstream performance.
    \item Although CoT reasoning introduces guided steps for query generation, this can sometimes result in diminished performance. CoT prompts may lead the model to generalize in a way that detracts from the relevance of the generated queries. For instance, in some cases, the reasoning steps cause the model to focus on peripheral concepts rather than extracting keywords or core topics directly relevant to the document.
\end{itemize}

These results, however, relate to the specific the datasets we test on. Further investigation across a larger pool of corpora is still needed. To answer then the question whether a filtering strategy is needed, the answer seems not evident as of now. Nonetheless, we note that using dynamic prompts instead of a static hardcoded template does improves the downstream performance of the reranker model, as indicated in the top scores for the 4 datasets shown.

\begin{table}[htbp]
    \small
    \centering
    \caption{Sample document from Scifact with its reference query (finding) and queries from the generator models. }
    \label{tab:query_eval}
    \begin{tabular}{m{0.15\linewidth}m{0.6\linewidth}m{0.08\linewidth}}
    \toprule
    & \textbf{Text} & \textbf{Score} \\
    \midrule
    \textbf{Document} & Myelodysplastic syndromes (MDS) are age-dependent stem cell malignancies that... & 100 \\
    \midrule
    \textbf{Reference} \textit{Finding} & Toll-like receptor (TLR) signaling is involved in the pathogenesis of human MDS. & 36.1 \\
    \textbf{Teacher} \textit{Finding} & Myeloid-derived suppressor cells (MDSC) are... hematopoiesis. MSDC expansion... &96.3\\
    \textbf{Student} \textit{Finding}& Myeloid-derived suppressor cells (MDSC) are ... of ineffective hematopoiesis. & 94.2\\ 
    \textbf{CPO} \quad \textit{Finding} & Myelodysplastic syndromes (MDS) are age-dependent stem cell malignancies that... & \textbf{99.6} \\
    \bottomrule
    \end{tabular}
\end{table}

\section{Conclusion}


In this study, we reproduce and verify the main claims of InPars Toolkit, confirming its claims of providing an end-to-end reproducible pipeline and a clear guide for reproducing InPars-V1, InPars-V2, and partially Promptagator. We also validate the pipeline’s plug-and-play capability by integrating LLaMA 3.1 8B as the generator model. Our results closely match the ones reported by the original authors, supporting the toolkit’s reliability.

Beyond reproduction, we extend the toolkit in two ways. First, we apply Contrastive Preference Optimization (CPO) to fine-tune the generator LLM to guide the model into generating generally better queries. 
Second, we apply CoT for dynamic prompt optimization in lieu of fixed templates. This generally improved query quality, though some cases revealed complexities which deserve further investigation.

Overall, our reproductions and extension experiments build upon the versatility of the InPars Toolkit and highlight practical pathways for future research.



\printbibliography
\appendix 

\end{document}